\documentclass[twocolumn,showpacs,preprintnumbers,amsmath,amssymb]{revtex4}

\usepackage{graphicx}
\usepackage{dcolumn}
\usepackage{bm}
\usepackage[colorlinks]{hyperref}

\begin{document}


\title{Emergent rewirings for cascades 
on correlated networks}

\author{Yukio Hayashi}
\author{Toshiyuki Miyazaki}
\affiliation{%
Japan Advanced Institute of Science and Technology,\\
Ishikawa, 923-1292, Japan
}%


\date{\today}

\begin{abstract}
Recent studies of attacks on complex networks suggest that
small initial breakdowns can lead to 
global cascades of overload failures in
 communication, economic trading, and supply-transportation systems,
considering the defense methods is very important to reduce the huge
damage.
The propagation of failures depends on the flow dynamics 
such as packet routing of physical quantities and
 on the heterogeneous distribution of load or capacity 
in many realistic networks which have topological 
properties of scale-free (SF) and 
degree-degree correlations.
We introduce a defense strategy based on emergent rewirings between 
 neighbors of the attacked node, 
and investigate the size of cascade on SF networks with 
controllable correlations.
We show the differences of damaged size for the types of correlations 
and the effective range of tolerance in the defense methods. 
They can be performed 
on the current 
wireless communication or dynamic allocation technologies.

\end{abstract}

\maketitle


Cascades of overload failures triggered by small initial failures or
attacks are sometimes occurred and propagated to very large damage 
in real networks such as power grid (blackout), 
Internet (packet congestion or 
server down by DoS), economic trading (bankruptcy), traffic system 
(jamming), and so forth.
This phenomenon is more severe than the disconnecting of networks 
by intentional attacks \cite{Albert00a}, 
because a physical quantity (hereafter called packet) can not be 
transmitted even through the connection if one of the 
terminal nodes exceeds its load capacity.
In other words, 
not only the topological structure of network 
but also the heterogeneously distributed load or capacity 
is deeply related to the intrinsic dynamics of packet flow
and to the size of cascade as the consequence of 
propagation of overload failures.

As an important topological structure,
we have found scale-free (SF) property 
in many complex networks \cite{Albert02}.
The highly heterogeneous degree distribution follows a power law
$P(k) \sim k^{- \gamma}$, $2 < \gamma < 3$.
Moreover, 
recent studies classify SF networks according to types of 
degree-degree
correlations of nodes with their neighbors \cite{Vazquez02};
social networks tend to have assortative connections between peers with
similar degrees \cite{Capocci03, Newman02c}, 
while technological or biological networks tend to have disassortative ones 
between those nodes with high degrees, namely hubs, and those with
low degrees \cite{Vazquez02, Newman03}.
The effect of correlations on cascades of overload failures 
is not yet studied, 
although a number of important aspects of cascading failures in complex
networks have been discussed \cite{Holme02, Moreno02, Watts02,
Albert04}; in particular 
a simple model of cascades of overload failures has been introduced 
\cite{Motter02} with a proposal of 
defense strategy based on intentional removals (IR)
\cite{Motter04}.
The numerical results 
have shown to reduce the size of cascade by the IR 
on a sacrifice of the removing a fraction of 
nodes that mainly contribute to generate
load rather than to transmit the packets.

We investigate the cascading failures on SF networks with 
degree-degree correlations,
and propose an alternative defense strategy based on 
emergent rewiring which is 
realized in distributed manner with additional cost in the neighbors of 
the initial attack.
On the current technologies, 
our proposed methods are basically applicable to wireless communications, 
airline flights, logistic networks, and so on 
in the practically important service domains.

The organization of this paper is as follows.
We first describe the basic process of cascading failures,
and consider the defense strategies.
Then, we introduce SF network models whose types of correlations are
controllable.
Through numerical simulations, we show the differences of damaged size 
for the types of correlations and the effect of our proposed 
defense strategy.
Finally, we summarize these results.

For a given (undirected) network, we assume that at each time step
a request of communication 
is generated between every pair of nodes $(i, j)$ 
and a packet is transmitted along the shortest paths connecting
nodes $i$ and $j$.
If there is more than one shortest paths connecting two given nodes,
the packet is divided evenly at each branching point.
In this situation, it is natural that the load $L_{k}(t)$ of node $k$
at time $t$ is defined through the betweenness centrality $B_{k}(t)$: 
the total amount of packets passing through the node 
per unit of time.
Note that the shortest paths may be changed by the propagation of failures.
Also as in Ref \cite{Motter02, Motter04},
the capacity $C_{k}$ of node $k$ is assumed to be proportional to its
initial load,
\begin{equation}
 C_{k} \stackrel{\rm def}{=} \alpha L_{k}(0), \;\;\;\;\;
 k = 1, 2, \ldots, N,
\end{equation}
where $\alpha \geq 1$ is the tolerance parameter, 
and $N$ is the initial number of nodes.
From  $\alpha \geq 1$, no node is overload in the initial network.
After an initial attack, the damaged node is disconnected.
Then, some nodes receive much loads that exceed own capacities 
by changing paths, 
and the corresponding nodes with overload collapse.
These removals lead to the next redistribution of load among the
remaining nodes in the network, 
and the subsequent overload may occur: 
the failures are propagated.
The cascading process is stop at $T$ with size $N'$ 
only when the updated load satisfies $L_{k}(T) \leq C_{k}$ 
for all the nodes $k$.
The damage caused by a cascade is quantified in terms of the relative
size of the largest connected component $GC = N' / N$.
As the minimum origin, 
we focus on 
load-based intentional attack upon an exhausted node
which has the highest load,
since global cascade can be triggered by removing 
only the key node \cite{Motter02}.

\begin{figure}[htb]
 \begin{center}
  \includegraphics[height=50mm]{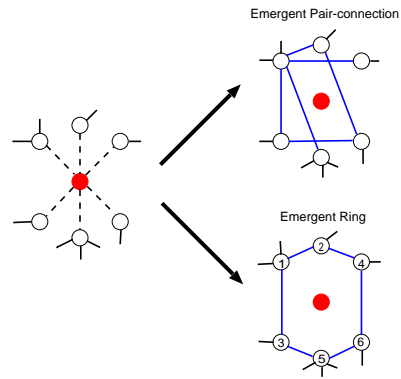}
  \caption{Defense strategy based on emergent rewirings.
  The red circle and dashed lines are the initially attacked node and
  the removed connections, respectively. Blue lines are the rewirings 
  between the neighbors. 
  The number in each circle denotes the order of $C_{i} / k_{i}$.} 
  \label{fig_adapt_rewire}
  \end{center}
\end{figure}

In addition to  
the defense strategy based on IR \cite{Motter04},
we introduce an alternative strategy based on 
emergent pair-connections (EP) and emergent ring (ER).
The $n$ connections from the initially attacked node are replaced by 
the rewirings between the neighbors
as shown in Fig. \ref{fig_adapt_rewire}.
Each of these strategies is performed after the initial attack before
the propagation.

\begin{description}
  \item[IR: ] The fraction $f$ of nodes with smallest 
	     $\Delta_{i} \stackrel{\rm def}{=} L_{i}(0) - L_{i}^{g}$
	     are removed to avoid the generation of packets from the
	     peripheral nodes, 
	     where the total load generated by node $i$ is 
	     \begin{equation}
	     L_{i}^{g} \stackrel{\rm def}{=} \sum_{j}
	     (D_{ij} + 1) = (\bar{D}_{i} + 1)(N - 1),
	     \end{equation}
	     $\bar{D}_{i} = \sum_{j} D_{ij} / N$,
	     $D_{ij}$ is the shortest path length 
	     between nodes $i$ and $j$.
  \item[EP: ] In the neighbors of the initial attack,
	     pairs of nodes are linked according to the
	     decreasing order of
	     \begin{equation}
	     W_{ij} \stackrel{\rm def}{=} C_{i} / k_{i} 
	     + C_{j} / k_{j}, 
	     \end{equation}
	     where $k_{i}$ and $k_{j}$ denotes the degrees of nodes 
	     $i$ and $j$.
	     Large $C_{i} / k_{i}$ corresponding to large 
	     $B_{i}(0)$ and small $k_{i}$ means that 
	     the node $i$ is a bridge node between subgraphs 
	     and that important to construct bypass routes.
  \item[ER: ] In the neighbors, a ring is rewired according to 
	     the decreasing order of $C_{i} / k_{i}$ as shown in 
	     Fig. \ref{fig_adapt_rewire} (right).
\end{description}
The variations of IR for selecting removed nodes 
have almost the same effect \cite{Motter04} 
because the quantities of smallest closeness centrality, load, 
and degree have correlations to smallest $\Delta_{i}$.
Similar modifications of EP and ER are also possible 
as mentioned latter.
Here we should remark that 
the conventional IR is cost-less, but needs the global information 
to chose the nodes with smallest $\Delta_{i}$ in the network;
while our proposed EP and ER modify local connections by using 
the capacities and degrees of the neighbors 
within only two steps, but not cost-less.

Next, we consider SF networks with various types of degree-degree 
correlations.
Unfortunately, 
we have no general approach to construct SF networks with a given
correlations such as the configuration method \cite{Newman01}
of random networks for a given degree distribution, 
in addition there are only a few models to be able to control the
correlations between assortative (Ass) and disassortative (Dis) networks.
Thus, we consider the following two models.
The procedures 
are repeated until the network reaches to the required size $N$.

CDD: coupled duplication-divergence model \cite{Vazquez03b}
\begin{enumerate}
  \item At each time step, 
	a new node $i'$ is added.
  \item Simultaneously, a node $i$
	is randomly chosen, and new connections 
	between all the neighbors $j$ of $i$ 
	and the new node $i'$ are duplicated.
  \item With probability $q_{v}$,
	a connection between $i$ and $i'$ is established
	(self-interaction).
  \item In the divergence process, 
	each duplicated connection is removed 
	with probability $1 - q_{e}$.
\end{enumerate}
These local rules are biologically plausible \cite{Sole02} and also
suitable for distributed systems.
Note that larger $q_{v}$ enhances the assortativity of 
network generated by the above rules 
because the self-interaction means connecting a pair of 
nodes with similar degrees.
In other words, 
$q_{v}$ is a control parameter of the correlation.
However, the number of total connections may vary because of the 
duplication processes.
Obviously, a network with many bypass routes as the consequence of many
connections 
becomes more tolerant to failures or attacks.
Thus, to match the condition, 
we confirm the variance is bounded in the 
self-averaging \cite{Ispolatov04} 
\begin{equation}
 \chi \stackrel{\rm def}{=} \frac{\sqrt{\langle k^{2} \rangle 
- \langle k \rangle^{2}}}{\langle k \rangle} < 0.2,
\end{equation}
for the set of parameters in Table \ref{table_param}.
The randomly generated CDD networks 
have $\langle k \rangle \approx 4$.

\begin{figure}
  \begin{minipage}[htb]{.47\textwidth}
    \includegraphics[height=50mm]{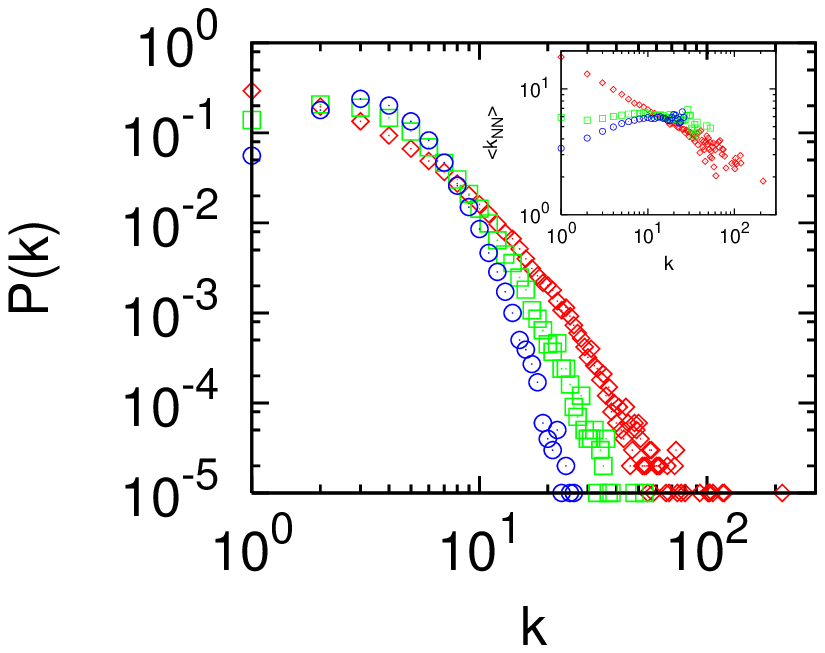}
    \begin{center} (a) CDD \end{center}
  \end{minipage} 
  \hfill 
  \begin{minipage}[htb]{.47\textwidth}
    \includegraphics[height=50mm]{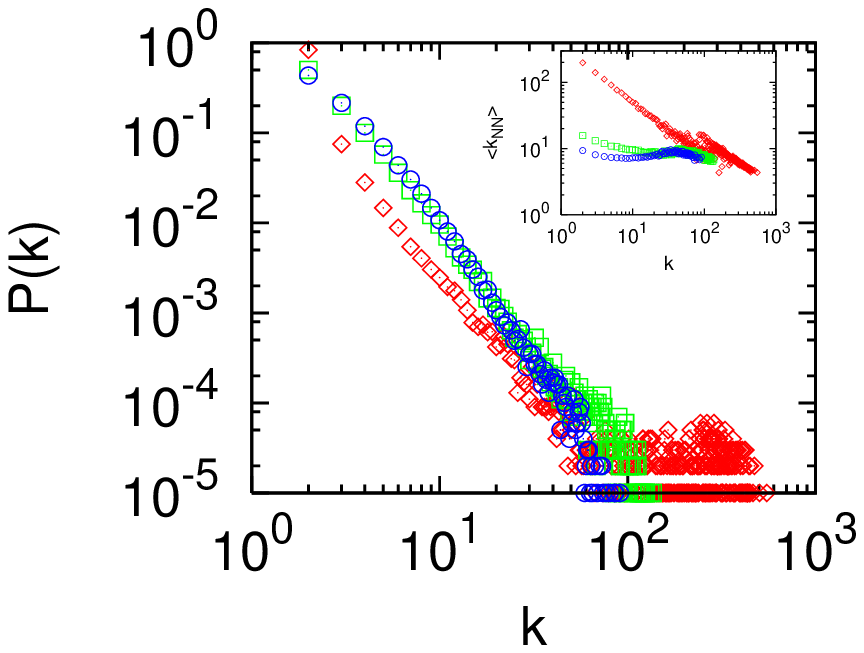} 
    \begin{center} (b) LPA \end{center}
  \end{minipage} 
  \caption{Degree distributions $P(k)$
 and degree-degree correlations $\langle k_{NN} \rangle$
 in the inset for (a) CDD and (b) LPA.
 The blue, green, and red marks correspond to 
 Ass, Unc, and Dis at the set of parameters in Table \ref{table_param}.
 They are the averages over 100 realizations.}
  \label{fig_deg_correl}
\end{figure}

LPA: shifted linear preferential attachment model
\cite{Barrat05}
\begin{enumerate}
  \item At each time step, 
	a new node is added and linked to old nodes
	by $m$ new connections.
  \item The attached nodes are randomly 
	chosen by the shifted linear preference 
	\cite{Barrat05, Krapivsky01}:
	a node with degree $k$ is chosen as the terminal of a new
	connection with probability proportional to $k + w$.
\end{enumerate}
This is a modification of the BA model \cite{Barabasi99} 
including it at $w = 0$ as uncorrelated networks (Unc) with 
$\langle k \rangle = 2 m$.
We set the parameter $w$ as in Table \ref{table_param} and $m = 2$
to satisfy $|w| < k_{min} = m$, 
where $k_{min}$ denotes the minimum degree.
The assortativity $r$
in Ref. \cite{Newman03} is calculated for each type of 
correlations.
Figs. \ref{fig_deg_correl}(a)(b) show the 
degree distributions follow power laws
with slightly collapsed parts; 
CDD restricts very low degrees because of the duplication process, 
and LPA with Dis have a star-like structure with a few hubs of very large
degrees.
The insets show the three types of correlations: Ass with a positive slope, 
Unc as almost flat, and Dis with a negative slope.

\begin{table}[htb]
\begin{center}
\begin{footnotesize}
\begin{tabular}{c||cc|c|c} \hline
     & CDD &             & LPA & assortativity \\\hline
Type & $q_{v}$ & $q_{e}$ & $w$   & $r$ \\ \hline
Ass & 1.0 & 0.26 & & 0.19\\ 
Unc & 0.5 & 0.35 & & 0.02\\ 
Dis & 0.0 & 0.42 & & -0.29\\ \hline
Weak Ass & &     & 1.8  & -0.01\\
Unc & &     & 0.0  & -0.08\\ 
Dis & &     & -1.8 & -0.49\\ \hline
\end{tabular}
\vspace{2mm}
\caption{A set of parameters for CDD and LPA.
The types of correlations Ass, Unc, and Dis denotes 
assortative, uncorrelated, and disassortative networks, respectively.
Unnecessary parameters are blanked.
The assortativity $r$ is 
measured over the 100 realizations of each network model with 
$N = 1,000$ and $\langle k \rangle \approx 4$.
}
\label{table_param}
\end{footnotesize}
\end{center}
\end{table}

In this setting, 
we numerically investigate the size of cascades and the effect of
defense strategies IR, EP, and ER for the SF networks.
Since the removal of the most central nodes can trigger global cascades
into a drastically reduced size, 
the defense strategies are based on 
the conventional IR: 
the restriction of load generation by removing a fraction $f$ of
nodes with most negative $\Delta_{i}$, 
and on the proposed EP and ER: 
emergent rewiring to compensate the bypass 
between bridge nodes with large 
$C_{i} / k_{i}$ (with relatively high centrality).

\begin{figure}
  \begin{minipage}[htb]{.47\textwidth}
    \includegraphics[height=60mm, angle=0]{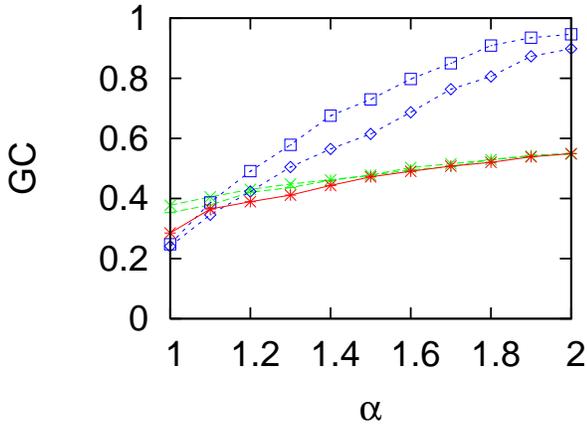} 
    \begin{center} (a) Ass \end{center}
  \end{minipage} 
  \hfill 
  \begin{minipage}[htb]{.47\textwidth}
    \includegraphics[height=60mm, angle=0]{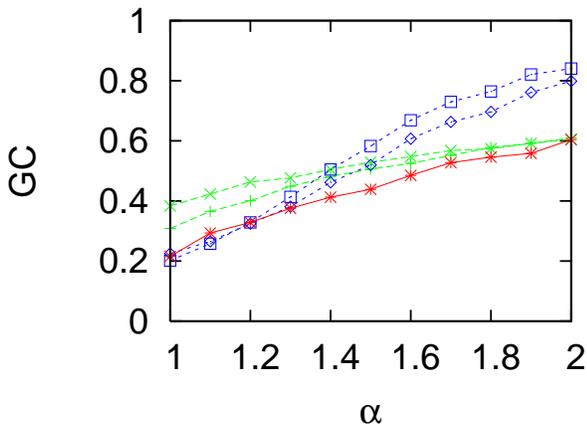} 
    \begin{center} (b) Unc \end{center}
  \end{minipage} 
  \hfill 
  \begin{minipage}[htb]{.47\textwidth} 
    \includegraphics[height=60mm, angle=0]{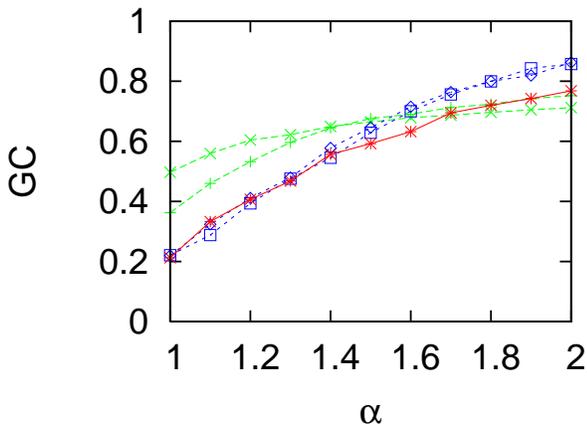} 
    \begin{center} (c) Dis \end{center}
  \end{minipage} 
\caption{Ratio GC as a function of the tolerance parameter $\alpha$
 for CDD with (a) Ass, (b) Unc, and  (c) Dis. 
 The solid red line marked by asterisk corresponds to NO: no defense. 
 The dashed green lines correspond to the conventional IR with 
 $f = 0.1$ (plus) and $f = 0.2$ (cross).
 The dashed blue lines correspond to EP (rectangle) and ER (diamond).
 They are the averages over 100 realizations.} 
\label{fig_comp_CDD}
\end{figure}

Figs. \ref{fig_comp_CDD} show the ratio GC as a function of the
tolerance parameter $\alpha$ for CDD.
The parts of dashed lines above the red line for 
no defense (NO) show that 
the corresponding strategies are effective.
We first compare the effect in the types of correlations,
and then discuss the difference in the defense strategies.
The damage caused by the intrinsic cascades with NO 
is larger as the correlation is Ass as shown 
in Fig. \ref{fig_comp_CDD}(a)
rather than Dis as shown in Fig. \ref{fig_comp_CDD}(c).
For the damage,
the IR with the nearly best fraction $f = 0.1$ or $0.2$ 
\cite{Motter04} remains a larger GC for Dis, 
while there is no such difference for the types of correlations 
in the EP and ER.
For the tolerance parameter, 
the effective range is contrast: 
small $\alpha < 1.5$ in the IR, however large $\alpha > 1.5$ 
in the EP and ER.
It is common for all types of correlations.
In particular, the size GC in the IR are strongly saturated 
for increasing $\alpha$, 
while that in the EP or ER is significantly larger.
For example, Fig. \ref{fig_comp_CDD}(a) shows 
the ratio GC is nearly 0.4 in the IR (or NO),
while it is improved to 0.6 $\sim$ 0.8 in the EP or ER 
at $\alpha = 1.5$.

\begin{figure}
  \begin{minipage}[htb]{.47\textwidth}
    \includegraphics[height=60mm, angle=0]{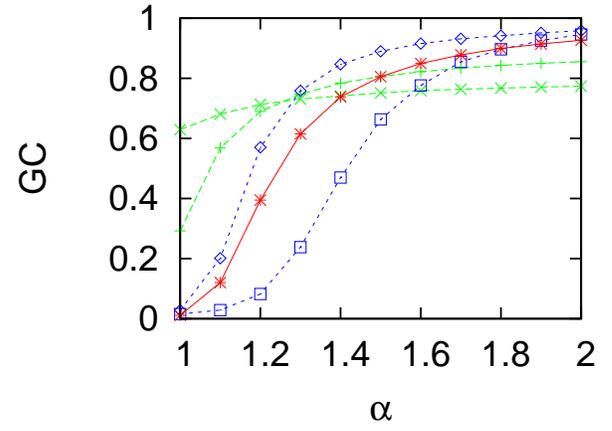} 
    \begin{center} (a) Ass \end{center}
  \end{minipage} 
  \hfill 
  \begin{minipage}[htb]{.47\textwidth}
    \includegraphics[height=60mm, angle=0]{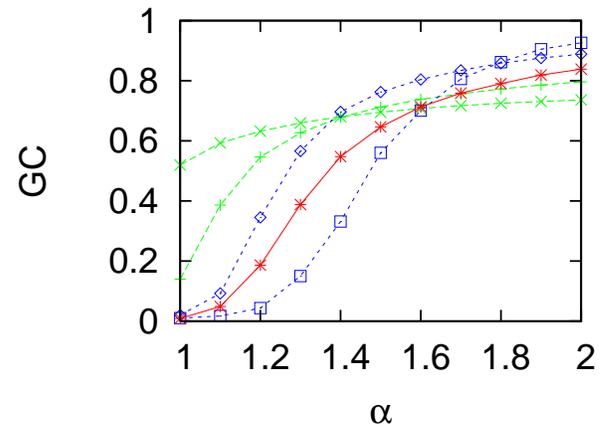} 
    \begin{center} (b) Unc \end{center}
  \end{minipage} 
  \hfill 
  \begin{minipage}[htb]{.47\textwidth} 
    \includegraphics[height=60mm, angle=0]{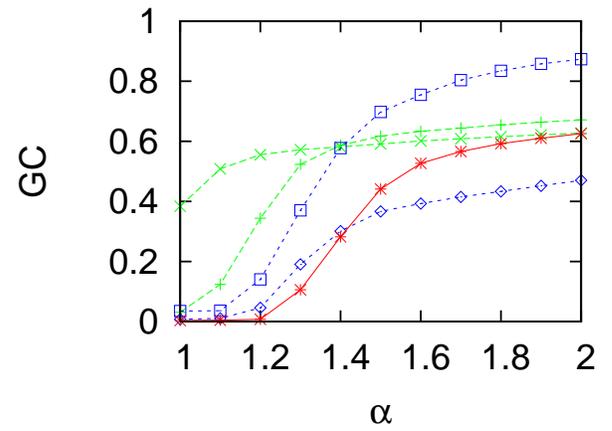} 
    \begin{center} (c) Dis \end{center}
  \end{minipage} 
\caption{Ratio GC as a function of the tolerance parameter $\alpha$
 for LPA with (a) Ass, (b) Unc, and  (c) Dis. 
 The lines and marks correspond to the same as the previous.
 They are the averages over 100 realizations.} 
\label{fig_comp_LPA}
\end{figure}

Figs. \ref{fig_comp_LPA} show the ratio GC as a function of the
tolerance parameter $\alpha$ for LPA.
In contrast to the results for CDD, 
the damage with NO is larger as the correlation is Dis 
as shown in Fig. \ref{fig_comp_LPA}(c) rather than Ass 
as shown in Fig. \ref{fig_comp_LPA}(a).
In particular, it is annihilative for small $\alpha < 1.2$.
Thus, LPA is very vulnerable, 
we have demonstrated \cite{Matsukubo05}
there is no effect in the modified EP and
ER by applying load or degree instead of $C_{i} / k_{i}$.
For the damage, the IR remains a larger GC for Ass
(but almost no effect by the saturation in $\alpha > 1.6$).
While the ER (marked by diamond) 
reduces the damage for Ass and Unc,
the EP (marked by rectangle) do for Dis in all the range 
$1 \leq \alpha \leq 2$.
On the other hand, as similar to the results for CDD, 
the IR is effective in small $\alpha< 1.3$,
but the EP or ER is in large $\alpha > 1.4$.

In summary, we have investigated 
the cascades of overload failures triggered from a load-based attack
in SF networks.
Such phenomena is widely observed 
in social and technological SF networks 
with degree-degree correlations.
We have shown that 
the size of cascades can be reduced by our defense strategy based on 
emergent rewirings between bridge nodes with large $C_{i} / k_{i}$,
and that 
the effect is slightly different for the types of correlations.
With the tolerance parameter $\alpha$,
they are more complicated: 
the conventional IR works better in tight capacity ($\alpha < 1.5$
for CDD with Dis and for LPA),
but the proposed 
EP or ER do in reasonable capacity ($\alpha > 1.5$ for CDD with
Ass and LPA with Dis).
We have confirmed 
similar results \cite{Miyazaki05} for the capacity in the 
other SF network models 
based on the connecting nearest neighbors \cite{Vazquez03b}
and on the triangulation \cite{Doye05}. 
Thus, rewiring is not always best,
however the effect is remarkable 
 in the reasonable capacity.
The ad hoc rewirings 
will be useful for protecting and sustaining our 
social organizations or 
communication infrastructures in real systems
from huge damage
triggered by small attacks against the vulnerable parts.

\end{document}